\documentclass[twocolumn,superscriptaddress,letter]{revtex4}%
\usepackage{amssymb}
\usepackage{color}
\usepackage{graphicx}
\usepackage{dcolumn}
\usepackage{bm}
\usepackage[header,title,page,titletoc]{appendix}
\usepackage{amsmath}
\usepackage{amsfonts}%
\providecommand{\U}[1]{\protect \rule{.1in}{.1in}}
﻿%2multibyte Version: 5.50.0.2890 CodePage: 936
﻿%2multibyte Version: 5.50.0.2890 CodePage: 936

\begin{document}
\title{General Exceptional Points}
\author{Xiao-Ran Wang}
\affiliation{Center for Advanced Quantum Studies, Department of Physics, Beijing Normal
University, Beijing 100875, China}
\author{Fei Yang}
\affiliation{Center for Advanced Quantum Studies, Department of Physics, Beijing Normal
University, Beijing 100875, China}
\author{Xiao-Jie Yu}
\affiliation{Center for Advanced Quantum Studies, Department of Physics, Beijing Normal
University, Beijing 100875, China}
\author{Xian-Qi Tong}
\affiliation{Center for Advanced Quantum Studies, Department of Physics, Beijing Normal
University, Beijing 100875, China}
\author{Su-Peng Kou}
\thanks{Corresponding author}
\email{spkou@bnu.edu.cn}
\affiliation{Center for Advanced Quantum Studies, Department of Physics, Beijing Normal
University, Beijing 100875, China}

\begin{abstract}
Exceptional points are interesting physical phenomena in non-Hermitian physics
at which the eigenvalues are degenerate and the eigenvectors coalesce. In this
paper, we find that the universal feature of arbitrary non-Hermitian two-level
systems with singularities is basis defectiveness rather than energy
degeneracy or state coalescence. This leads to the discovery of general
exceptional points (GEPs).\ For GEPs, more subtle structures (e.g., the
so-called Bloch peach), additional classification, and \textquotedblleft
hidden\textquotedblright \ quantum phase transitions are explored. By using the
topologically protected sub-space from two edge states in the non-Hermitian
Su--Schrieffer--Heeger model as an example, we illustrate the physical
properties of different types of GEPs.

\end{abstract}
\maketitle

\section{Introduction}

\textquotedblleft \emph{Exceptional point}\textquotedblright \ (EP) is a
mathematical term introduced by T. Kato over half a century ago\cite{Kato}. In
mathematics, EPs are branch point singularities of a spectrum and
eigenfunctions for non-Hermitian matrices\cite{Heiss1,Moiseyev}. In the form
of Jordan block matrix, at EPs the algebra multiplier of a matrix becomes
larger than its geometric multiplier. Since the publication of the related
paper by Bender and Boettcher\cite{Bender1}, EPs have become one of the most
interesting phenomena in non-Hermitian physics\cite{Miri,Minganti}. As a
particular example, for non-Hermitian Hamiltonians with parity--time
($\mathcal{PT}$)
symmetry\cite{Bender2,Heiss2,Mostafazadeh,Lee,Kawabata,El-Ganainy,Yao,Rotter,Jin,Deng,Liu,Ran,Ashida,Guo,Guo1,wang,wang1}%
, spontaneous $\mathcal{PT}$-symmetry breaking corresponds to a typical EP, at
which the energy levels become degenerate and the eigenvectors coalesce. In
experiments, the phenomenon of EPs has been realized and simulated using
various
approaches\cite{Feng,Peng,Hodaei,AGuo,Chen,Hodaei1,El-Ganainy2,Schindler,Feng2,Regensburger,Bittner,MBender,Hang,Zhu,Brandstetter,Popa,Fleury,Zhen,Xu,Peng1,Zhang,Assawaworrarit,Choi,Scheel,Wu,L.
Xiao1,Naghiloo,L. Xiao2,Liu1,ChenPX,Ding}.

On the other hand, in some quantum many-body models, due to special conditions
of symmetry/topology, there may exist protected sub-systems. For example, for
topological insulators, there exist topologically protected edge states with
gapless energy spectra (or zero modes for one dimensional
cases)\cite{Kane2010,Qi2011}; for the many-body systems with spontaneously
symmetry breaking there exist symmetry-protected degenerate ground states; for
the topological orders with long range entanglement, there exist topologically
protected degenerate ground states (on a torus) that make up topological
qubits and{ may be possible to be applied to incorporate intrinsic fault
tolerance into a quantum computer}\cite{Kitaev,wen,wen1,S. P. Kou}.

For these topologically/symmetry protected sub-systems in different quantum
many-body models, the quantum properties will be changed under non-Hermitian
perturbations. \emph{How the non-Hermitian perturbations affect the
topologically/symmetry protected sub-systems? Do there exist new types of
EPs?} In this paper, to answer these questions, we investigate the properties
of EPs and find that basis defectiveness plays a key role in EPs. Furthermore,
we find that in certain non-Hermitian systems (subsystems of certain
non-Hermitian models) there may exist EPs without eigenvalue degeneracy, EPs
without the coalescence of different eigenvectors, or those without both
features (see the following discussions). This leads to the discovery of
general EPs.

The remainder of the paper is organized as follows. In Sec. II, we review the
theory of singularity for usual EPs in a simple two-level systems and show the
reason why eigenstates coalesce at an EP. In Sec. III, we discuss a general
theory for protected non-Hermitian sub-spaces (or topologically/symmetry
protected sub-systems in different quantum many-body models) and show how to
derive the effective Hamiltonian and the corresponding (initial) basis. In
Sec. IV, we develop a theory of singularity for the general EPs and show their
complete classification. In Sec. V, an example of a one-dimensional (1D)
non-Hermitian topological insulator -- 1D nonreciprocal Su--Schrieffer--Heeger
(SSH) model is focused on. According to the global phase diagram of the
topologically protected sub-systems from the two edge states, we show the
occurrence of different types of general EPs. Finally, the conclusions are
given in Sec. VI.

\section{Singularity at an EP in a non-Hermitian two-level $\mathcal{PT}$
system}

\subsection{EP in a non-Hermitian two-level $\mathcal{PT}$ system}

To learn the nature of the singularity at an EP, we study a two-level
$\mathcal{PT}$ system described by the following Hamiltonian:
\begin{equation}
\hat{H}_{\mathrm{NH}}=h_{x}\sigma_{x}+ih_{z}\sigma_{z}.
\end{equation}
For this two-level non-Hermitian system, $\{|\psi_{0}^{\mathrm{R}}%
\rangle \}=\{|\psi_{0}\rangle \}=\{ \left \vert \uparrow \right \rangle ,\left \vert
\downarrow \right \rangle \}$ denotes an \emph{initial basis} obeying orthogonal
and normalization conditions, i.e., $\langle \uparrow|\downarrow \rangle=0$ and
$\langle \uparrow|\uparrow \rangle=\langle \downarrow|\downarrow \rangle=1$.

At $h_{x}=h_{z}$, a typical spontaneous $\mathcal{PT}$-symmetry breaking
occurs: for the case of $h_{x}>h_{z},$ the energy levels $\left \vert
+\right \rangle $ and $\left \vert -\right \rangle $\textit{ }are $E_{\pm}%
=\pm \sqrt{h_{x}^{2}-h_{z}^{2}};$ For the case $h_{x}<h_{z},$ these two energy
levels are\textit{ }$E_{\pm}=\pm i\sqrt{h_{z}^{2}-h_{x}^{2}};$ For the case of
$h_{x}=h_{z}$, the system is at an EP with eigenstate coalescence and energy
degeneracy. To strictly characterize the coalescence of the eigenstates, we
define their state similarity, $\Lambda=|\langle \tilde{\psi}_{+}^{\mathrm{R}%
}|\tilde{\psi}_{-}^{\mathrm{R}}\rangle|$, where $|\tilde{\psi}_{+}%
^{\mathrm{R}}\rangle$ and $|\tilde{\psi}_{-}^{\mathrm{R}}\rangle$ are
eigenstates satisfying the self-normalization conditions $|\langle \tilde{\psi
}_{+}^{\mathrm{R}}|\tilde{\psi}_{+}^{\mathrm{R}}\rangle|=1$ and $|\langle
\tilde{\psi}_{-}^{\mathrm{R}}|\tilde{\psi}_{-}^{\mathrm{R}}\rangle|=1$. Fig.
1(a) and Fig. 1(b) show the two energy levels and the state similarity
$\Lambda$, respectively. It is obvious that both $E_{+}=E_{-}$ and $\Lambda=1$
hold at the EP ($h_{x}=h_{z}$).

\begin{figure}[ptb]
\includegraphics[clip,width=0.62\textwidth]{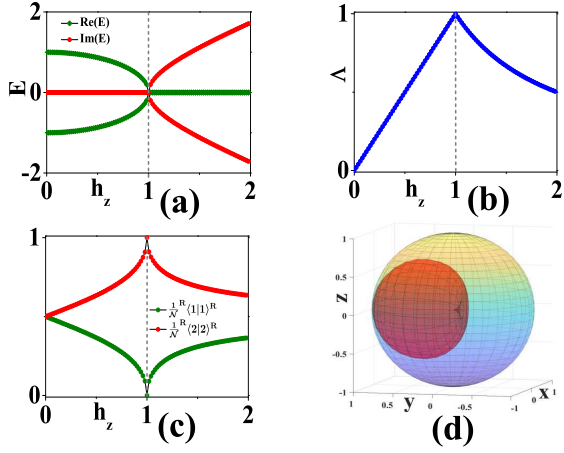} \centering
\setlength{\abovecaptionskip}{-1.cm} \vspace{-0.2cm}\caption{(Color online)
Physical properties of the EP ($h_{x}=h_{z}$) for the non-Hermitian two-level
model $\hat{H}_{\mathrm{NH}}=h_{x}\sigma_{x}+ih_{z}\sigma_{z}$. $h_{x}$ is set
to unity. (a) The two energy levels, $E_{\pm}.$ At the EP ($h_{x}=h_{z}$),
these two energy levels become degenerate; (b) The state similarity
$\Lambda=|\langle \tilde{\psi}_{+}^{\mathrm{R}}|\tilde{\psi}_{-}^{\mathrm{R}%
}\rangle|$ for two eigenstates $|\tilde{\psi}_{+}^{\mathrm{R}}\rangle$ and
$|\tilde{\psi}_{-}^{\mathrm{R}}\rangle$ satisfying the self-normalization
conditions $|\langle \tilde{\psi}_{+}^{\mathrm{R}}|\tilde{\psi}_{+}%
^{\mathrm{R}}\rangle|=1$ and $|\langle \tilde{\psi}_{-}^{\mathrm{R}}%
|\tilde{\psi}_{-}^{\mathrm{R}}\rangle|=1$. At the EP ($h_{x}=h_{z}$), we have
$\Lambda=1$; (c) The self-normalization of the matrix basis, $\frac
{1}{\mathcal{N}}^{\mathrm{R}}\langle1|1\rangle^{\mathrm{R}}$ and $\frac
{1}{\mathcal{N}}^{\mathrm{R}}\langle2|2\rangle^{\mathrm{R}}$, where
$\mathcal{N}=^{\mathrm{R}}\langle1|1\rangle^{\mathrm{R}}+^{\mathrm{R}}%
\langle2|2\rangle^{\mathrm{R}}.$ At the EP ($h_{x}=h_{z}$), we have $\frac
{1}{\mathcal{N}}^{\mathrm{R}}\langle1|1\rangle^{\mathrm{R}}=1$ and $\frac
{1}{\mathcal{N}}^{\mathrm{R}}\langle2|2\rangle^{\mathrm{R}}=0$; (d) The Bloch
peach for the quantum states $|\psi^{\mathrm{R}}\rangle=\cos \frac{\theta}%
{2}\left \vert 1\right \rangle ^{\mathrm{R}}+e^{i\phi}\sin \frac{\theta}%
{2}\left \vert 2\right \rangle ^{\mathrm{R}}$ at the EP ($h_{x}=h_{z}$). Here,
we have $^{\mathrm{R}}\langle1|2\rangle^{\mathrm{R}}=0$, $^{\mathrm{R}}%
\langle1|1\rangle^{\mathrm{R}}=1$, and $^{\mathrm{R}}\langle2|2\rangle
^{\mathrm{R}}=e^{-2\beta^{M}}$. }%
\end{figure}

\subsection{Singular non-Hermitian similar transformation}

Let us understand \emph{why eigenvectors coalesce at EPs. }

By performing a non-Hermitian similarity transformation $\mathcal{\hat{S}}%
_{M}=e^{-\beta^{M}\sigma_{y}}$, we transform the original non-Hermitian
Hamiltonian $\hat{H}_{\mathrm{NH}}$ into a Hermitian/anti-Hermitian one,
$\hat{H}_{0}$ ($\hat{H}_{0}^{\dagger}=\hat{H}_{0}$ or $\hat{H}_{0}^{\dagger
}=-\hat{H}_{0}$), i.e.,
\begin{equation}
\hat{H}_{\mathrm{NH}}\rightarrow \mathcal{\hat{S}}_{M}^{-1}\hat{H}%
_{\mathrm{NH}}\mathcal{\hat{S}}_{M}=\hat{H}_{0},
\end{equation}
where $\beta^{M}=\left \vert \frac{1}{2}\ln \left \vert \frac{h_{x}+h_{z}}%
{h_{x}-h_{z}}\right \vert \right \vert $. The eigenvalue of $\hat{H}_{0}$ is the
same as that of $\hat{H}_{\mathrm{NH}}.$ Under the non-Hermitian\ similarity
transformation $\mathcal{\hat{S}}_{M}$, the basis of $\hat{H}_{\mathrm{NH}}$
is correspondingly changed, i.e.,
\begin{equation}
\{|\psi_{0}^{\mathrm{R}}\rangle \} \rightarrow \{|\psi^{\mathrm{R}}\rangle \}=\{
\mathcal{\hat{S}}_{M}(\sigma_{y},\beta^{M})|\psi_{0}\rangle \}.
\end{equation}
In the following, we refer to $\{|\psi^{\mathrm{R}}\rangle \}$ as the
\emph{matrix basis}.\ If we select the initial basis $\{|\psi_{0}^{\mathrm{R}%
}\rangle \}$ to be the eigenstates of $\sigma_{y},$, i.e.,
\begin{align}
\{|\psi_{0}^{\mathrm{R}}\rangle \}  &  =\{ \left \vert 1\right \rangle
,\left \vert 2\right \rangle \} \nonumber \\
&  =\{ \frac{1}{\sqrt{2}}(\left \vert \uparrow \right \rangle +i\left \vert
\downarrow \right \rangle ),\frac{1}{\sqrt{2}}(\left \vert \uparrow \right \rangle
-i\left \vert \downarrow \right \rangle )\},
\end{align}
then the matrix basis $\{|\psi^{\mathrm{R}}\rangle \}$ becomes
\begin{equation}
\{|\psi^{\mathrm{R}}\rangle \}=\{e^{-\beta^{M}\sigma_{y}}|\psi_{0}\rangle \}=\{
\left \vert 1\right \rangle ^{\mathrm{R}},\left \vert 2\right \rangle
^{\mathrm{R}}\},
\end{equation}
where $\left \vert 1\right \rangle ^{\mathrm{R}}=\left \vert 1\right \rangle $ and
$\left \vert 2\right \rangle ^{\mathrm{R}}=e^{-\beta^{M}}\left \vert
2\right \rangle $. The matrix basis $\{ \left \vert 1\right \rangle ^{\mathrm{R}%
},\left \vert 2\right \rangle ^{\mathrm{R}}\}$ obeys the orthogonal condition
$^{\mathrm{R}}\langle1|2\rangle^{\mathrm{R}}=0$ but does not obey the
normalization conditions, i.e., $^{\mathrm{R}}\langle1|1\rangle^{\mathrm{R}%
}=1$ and $^{\mathrm{R}}\langle2|2\rangle^{\mathrm{R}}=e^{-2\beta^{M}}$.

Approaching the EP, the non-Hermitian similarity transformation becomes
singular, i.e.,
\begin{equation}
\mathcal{\hat{S}}_{M}(\sigma_{y},\beta^{M})=e^{-\beta^{M}\sigma_{y}}%
\end{equation}
with $\beta^{M}=\left \vert \frac{1}{2}\ln \left \vert \frac{h_{x}+h_{z}}%
{h_{x}-h_{z}}\right \vert \right \vert \rightarrow \infty$. As a result, the
matrix basis becomes defective, i.e.,
\begin{align}
\{|\psi^{\mathrm{R}}\rangle \}  &  =\{ \mathcal{\hat{S}}_{M}(\sigma_{y}%
,\beta^{M}\rightarrow \infty)|\psi_{0}\rangle \} \nonumber \\
&  =\{ \left \vert 1\right \rangle ^{\mathrm{R}},\left \vert 2\right \rangle
^{\mathrm{R}}\} \rightarrow \{ \left \vert 1\right \rangle ^{\mathrm{R}},0\}.
\end{align}
Here, the base $\left \vert 2\right \rangle ^{\mathrm{R}}$\  \emph{disappears}!
To illustrate the defectiveness of the matrix basis, in Fig. 1(c), we plot
$\frac{1}{\mathcal{N}}^{\mathrm{R}}\langle1|1\rangle^{\mathrm{R}}$ and
$\frac{1}{\mathcal{N}}^{\mathrm{R}}\langle2|2\rangle^{\mathrm{R}}$, where
$\mathcal{N}=^{\mathrm{R}}\langle1|1\rangle^{\mathrm{R}}+^{\mathrm{R}}%
\langle2|2\rangle^{\mathrm{R}}$\textit{ }is a normalization factor. Near the
EP, $\frac{1}{\mathcal{N}}^{\mathrm{R}}\langle2|2\rangle^{\mathrm{R}}$ becomes
zero ($^{\mathrm{R}}\langle2|2\rangle^{\mathrm{R}}\rightarrow0$).

As a result, according to the defective matrix basis without $\left \vert
2\right \rangle ^{\mathrm{R}},$ the two eigenstates $|\tilde{\psi}%
_{+}^{\mathrm{R}}\rangle$ and $|\tilde{\psi}_{-}^{\mathrm{R}}\rangle$\ must
coalesce into $\left \vert 1\right \rangle $, i.e.,
\begin{align}
|\tilde{\psi}_{+}^{\mathrm{R}}\rangle &  =\frac{1}{\sqrt{1+e^{-2\beta^{M}}}%
}(\left \vert 1\right \rangle ^{\mathrm{R}}+\left \vert 2\right \rangle
^{\mathrm{R}})\nonumber \\
&  \rightarrow \frac{1}{\sqrt{1+e^{-2\beta^{M}}}}\left \vert 1\right \rangle
\end{align}
and
\begin{align}
|\tilde{\psi}_{-}^{\mathrm{R}}\rangle &  =\frac{1}{\sqrt{1+e^{-2\beta^{M}}}%
}(\left \vert 1\right \rangle ^{\mathrm{R}}-\left \vert 2\right \rangle
^{\mathrm{R}})\nonumber \\
&  \rightarrow \frac{1}{\sqrt{1+e^{-2\beta^{M}}}}\left \vert 1\right \rangle .
\end{align}
Now, near the EP, where $\beta^{M}\rightarrow \infty$, the state similarity
$\Lambda=|\langle \tilde{\psi}_{+}^{\mathrm{R}}|\tilde{\psi}_{-}^{\mathrm{R}%
}\rangle|=\frac{1}{1+e^{-2\beta^{M}}}$ obviously approaches $1$.\ Thus, one
can see that the singularity of the EP arises from the defective matrix basis
$\{ \left \vert 1\right \rangle ^{\mathrm{R}},0\}$ due to\emph{ }the singular
non-Hermitian similarity transformation $\mathcal{\hat{S}}_{M}(\sigma
_{y},\beta^{M}\rightarrow \infty)$.

\subsection{Bloch peach for two-level states under the non-Hermitian
similarity transformation}

To further illustrate the singularity of EPs, we use a geometric approach to
illustrate the deformation of the Bloch sphere under a singular non-Hermitian
similarity transformation $\mathcal{\hat{S}}_{M}(\sigma_{y},\beta
^{M}\rightarrow \infty)$.

For the Hermitian case, one may use a point on the Bloch sphere with
\textrm{SU(2)} rotation symmetry to represent an arbitrary quantum state of
the two-level system, i.e.,
\begin{equation}
|\psi \rangle=\cos \frac{\theta}{2}\left \vert 1\right \rangle +e^{i\varphi}%
\sin \frac{\theta}{2}\left \vert 2\right \rangle .
\end{equation}
Here, $\theta \in \lbrack0,\pi]$ and $\varphi \in \lbrack0,2\pi]$ are real
numbers, and the radius $r$ of the Bloch sphere\ can be obtained as
$r=\langle \psi|\psi \rangle=1$.

For a two-level $\mathcal{PT}$ system, a quantum state under the non-Hermitian
similarity transformation $\mathcal{\hat{S}}_{M}(\sigma_{y},\beta^{M})$
becomes
\begin{equation}
|\psi^{\mathrm{R}}\rangle=\mathcal{\hat{S}}_{M}|\psi \rangle=\cos \frac{\theta
}{2}\left \vert 1\right \rangle +e^{i\varphi}e^{-\beta^{M}}\sin \frac{\theta}%
{2}\left \vert 2\right \rangle
\end{equation}
(with $\theta \in \lbrack0,\pi]$ and $\varphi \in \lbrack0,2\pi]$), and the radius
$R$ of the Bloch sphere\ can be obtained as
\begin{equation}
R=\langle \psi^{\mathrm{R}}|\psi^{\mathrm{R}}\rangle=\cos^{2}\frac{\theta}%
{2}+e^{-2\beta^{M}}\sin^{2}\frac{\theta}{2}.
\end{equation}

Therefore, the original Bloch sphere changes into a peach-like closed surface
with residue \textrm{U(1)} symmetry along the y-axis (we call this the
\emph{Bloch peach}). With increasing $\beta$, one pole of the Bloch peach
moves upward. At the EP, this pole touches the origin of the coordinate
system. See the illustration of the Bloch peach in the limit of $\beta
^{M}\rightarrow \infty$ in Fig. 1(d).

\section{Protected non-Hermitian sub-systems}

Before giving the definition of new type of EPs, we define \emph{protected
non-Hermitian sub-systems}.

Due to special conditions of symmetry/topology, there may exist protected
sub-systems. To completely characterize the many-body model and its protected
sub-systems, we give their definitions, $\{ \hat{H}_{\mathrm{MB}},$
$B_{\mathrm{MB}}\}$ and $\{ \hat{H}_{\mathrm{S}},$ $B_{\mathrm{S}}\},$
respectively. Here, both $\hat{H}_{\mathrm{MB}}$\ and $\hat{H}_{\mathrm{S}}%
$\ are Hermitian Hamiltonians, and both $B_{\mathrm{MB}}=\{|\Psi_{j}%
\rangle,j=1,2,...,N\}$ and $B_{\mathrm{S}}=\{ \mathcal{P}|\Psi_{j}%
\rangle,j=1,2,...,N\}=\{|\mathrm{S}_{j}\rangle,j=1,2,...,K\}$ are normal basis
with $B_{\mathrm{S}}=\mathcal{P}B_{\mathrm{MB}}\in B_{\mathrm{MB}}$ where
$\mathcal{P}$ is a projected operator on the basis of the many-body system. It
is obvious that $K<N$.

When one adds a non-Hermitian perturbation $i\delta \hat{H}$ on the many-body
model, the total Hamiltonian becomes non-Hermitian, i.e.,
\begin{equation}
\hat{H}_{\mathrm{MB}}\rightarrow \hat{H}_{\mathrm{NH-MB}}=\hat{H}_{\mathrm{MB}%
}+i\delta \hat{H}.
\end{equation}
Because the basis has no changing, the many-body model is denoted by $\{
\hat{H}_{\mathrm{NH-MB}},$ $B_{\mathrm{MB}}\}$. In general, for a
non-Hermitian sub-system, the biorthogonal set for the basis is defined by
$|\Psi_{j}^{\mathrm{R}}\rangle$ and $|\Psi_{j}^{\mathrm{L}}\rangle$
($j=1,2,...,K$), i.e.,
\begin{equation}
\hat{H}_{\mathrm{NH-MB}}|\Psi_{j}^{\mathrm{R}}\rangle=E_{j}|\Psi
_{j}^{\mathrm{R}}\rangle,
\end{equation}
and
\begin{equation}
\hat{H}_{\mathrm{NH-MB}}^{\dagger}|\Psi_{j}^{\mathrm{L}}\rangle=(E_{j})^{\ast
}|\Psi_{j}^{\mathrm{L}}\rangle,
\end{equation}
and $\langle \Psi_{j}^{\mathrm{L}}|\Psi_{j}^{\mathrm{R}}\rangle=1$ where $j$ is
state index.

We then assume the non-Hermitian terms are perturbation and don't change the
existence of the protected sub-system and use $\{ \hat{H}_{\mathrm{NH-S}%
},B_{\mathrm{S}}\}=\{ \mathcal{P}\hat{H}_{\mathrm{MB}}\mathcal{P}%
^{-1},\mathcal{P}B_{\mathrm{MB}}\}$ to describe the protected (non-Hermitian)
sub-system. Here, $\hat{H}_{\mathrm{NH-S}}=\mathcal{P}\hat{H}_{\mathrm{MB}%
}\mathcal{P}^{-1}$ is the effective Hamiltonian of the protected sub-system.
Under the projected operation $\mathcal{P},$ the basis of the protected
sub-system is obtained as
\begin{align}
B_{\mathrm{MB}}  &  =\{|\Psi_{j}^{\mathrm{R}}\rangle \} \\
&  \rightarrow B_{\mathrm{S}}=\mathcal{P}B_{\mathrm{MB}}=\{ \mathcal{P}%
|\Psi_{j}^{\mathrm{R}}\rangle \}=\{|\psi_{j}^{\mathrm{R}}\rangle \}.\nonumber
\end{align}

In addition, we show the method to calculate $\mathcal{\hat{H}}_{\mathrm{NH}%
}.$ Based on the basis $S_{\mathrm{NH}}=\{|\psi_{j}^{\mathrm{R}}\rangle \}$, an
effective Hamiltonian of protected sub-system is derived as
\begin{equation}
\hat{H}_{\mathrm{NH-S}}=%
%TCIMACRO{\dsum \limits_{ij}}%
%BeginExpansion
{\displaystyle \sum \limits_{ij}}
%EndExpansion
h_{IJ}%
\end{equation}
where
\begin{equation}
h_{IJ}=\left \langle \psi_{j}^{\mathrm{L}}\right \vert \hat{H}_{\mathrm{NH}%
}\left \vert \psi_{j}^{\mathrm{R}}\right \rangle ,\text{ }I,J=1,2,...,K.
\end{equation}
In particular, the basis of protected sub-sysrem $B_{\mathrm{S}}=\{|\psi
_{j}^{\mathrm{R}}\rangle \}$ is always \emph{abnormal.} For example, it does't
follow usual normalization $\langle \psi_{j}^{\mathrm{R}}|\psi_{j}^{\mathrm{R}%
}\rangle \neq1$. In this paper, we only focus on the case of $K=2$.

\section{General EPs -- universal feature and classification}

We next develop a universal theory for an arbitrary non-Hermitian protected
two-level sub-system with a singularity and introduce the concept of
\emph{general EPs}. This phenomenon always occurs in subsystems of certain
non-Hermitian models, for example, the defective edge states of a
non-Hermitian topological insulator, the defective degenerate ground states in
non-Hermitian systems with spontaneous symmetry breaking, or the topologically
protected degenerate ground states in intrinsic topological orders.

A general non-Hermitian protected two-level sub-system is described by $\{
\hat{H}_{\mathrm{NH}},\{|\psi_{0}^{\mathrm{R}}\rangle \} \}$. The Hamiltonian
is
\begin{equation}
\hat{H}_{\mathrm{NH}}=h_{0}+\vec{h}\cdot \vec{\sigma},
\end{equation}
where $h_{0}$ is a complex number, $\vec{h}=(h^{x},$ $h^{y},$ $h^{z})$ is a
complex vector and $\vec{\sigma}=(\hat{\sigma}_{x},\hat{\sigma}_{y}%
,\hat{\sigma}_{z})$ is the vector of Pauli matrices. $\{|\psi_{0}^{\mathrm{R}%
}\rangle \}=\{ \mathcal{\hat{S}}_{B}|\psi_{0}\rangle \}$ denotes the initial
basis where $\mathcal{\hat{S}}_{B}=e^{-\beta^{B}\cdot \vec{\sigma}^{B}}$ is a
non-Hermitian\ similarity transformation with a real positive $\beta^{B}$ and
$|\psi_{0}\rangle$ is a normal basis consisting of eigenstates of $\vec
{\sigma}^{B}$ (or $\vec{\sigma}^{B}|\psi_{0}\rangle=\pm|\psi_{0}\rangle$).

\subsection{Representation under matrix basis}

To identify the universal feature of general EPs, we transform the original
non-Hermitian Hamiltonian $\hat{H}_{\mathrm{NH}}$ into a
Hermitian/anti-Hermitian one and derive its representation under matrix basis,
$\{ \hat{H}_{0},\{|\psi^{\mathrm{R}}\rangle \} \}$.

First, we transform the non-Hermitian Hamiltonian $\hat{H}_{\mathrm{NH}}%
=h_{0}+\vec{h}\cdot \vec{\sigma}$ into
\begin{equation}
\hat{H}_{\mathrm{NH}}=h_{0}+(\operatorname{Re}\vec{h})\vec{\sigma
}_{\operatorname{Re}}+i(\operatorname{Im}\vec{h})\vec{\sigma}%
_{\operatorname{Im}},
\end{equation}
where $\vec{\sigma}_{\operatorname{Re}}$ and $\vec{\sigma}_{\operatorname{Im}%
}$ ($\vec{\sigma}_{\operatorname{Re}}\cdot \vec{\sigma}_{\operatorname{Re}}=1$
and $\vec{\sigma}_{\operatorname{Im}}\cdot \vec{\sigma}_{\operatorname{Im}}=1$)
are the Pauli matrices corresponding to the real and imaginary parts of
$\vec{h}$, respectively.

Second, we divide $i(\operatorname{Im}\vec{h})\vec{\sigma}_{\operatorname{Im}%
}$ into two parts, $i(\operatorname{Im}\vec{h})^{A}\vec{\sigma}%
_{\operatorname{Im}}^{A}$ and $i(\operatorname{Im}\vec{h})^{C}\vec{\sigma
}_{\operatorname{Im}}^{C}$, with $\left[  \vec{\sigma}_{\operatorname{Im}}%
^{C},\vec{\sigma}_{\operatorname{Re}}\right]  =0$ and $\{ \vec{\sigma
}_{\operatorname{Im}}^{A},\vec{\sigma}_{\operatorname{Re}}\}=0.$ Then, the
non-Hermitian Hamiltonian becomes
\begin{equation}
\hat{H}_{\mathrm{NH}}=h_{0}+\vec{h}_{\operatorname{Re}}^{\prime}\cdot
\vec{\sigma}_{\operatorname{Re}}+i\vec{h}_{\operatorname{Im}}^{\prime}%
\cdot \vec{\sigma}_{\operatorname{Im}}^{A},
\end{equation}
where $\vec{h}_{\operatorname{Re}}^{\prime}=(\operatorname{Re}\vec
{h})+i(\operatorname{Im}\vec{h})^{C}$ and $\vec{h}_{\operatorname{Im}}%
^{\prime}=(\operatorname{Im}\vec{h})^{A}$.

Thirdly, under a non-Hermitian\ similarity transformation $\mathcal{\hat{S}%
}_{M}(\vec{\sigma}^{M},\beta^{M})$, we transform the original NH Hamiltonian
$\hat{H}_{\mathrm{NH}}$ into a Hermitian/anti-Hermitian one, i.e.,
\begin{align}
\hat{H}_{\mathrm{NH}}  &  \rightarrow \mathcal{\hat{S}}_{M}^{-1}(\vec{\sigma
}^{M},\beta^{M})\hat{H}_{\mathrm{NH}}\mathcal{\hat{S}}_{M}(\vec{\sigma}%
^{M},\beta^{M})\nonumber \\
&  =\hat{H}_{0}=h_{0}+\sqrt{\vec{h}^{2}}\vec{\sigma}_{\operatorname{Re}},
\end{align}
where
\begin{equation}
\mathcal{\hat{S}}_{M}(\vec{\sigma}^{M},\beta^{M})=e^{-\beta^{M}\cdot
\vec{\sigma}^{M}}%
\end{equation}
with $\vec{\sigma}^{M}=\frac{1}{2i}[\vec{\sigma}_{\operatorname{Im}}^{A}%
,\vec{\sigma}_{\operatorname{Re}}]$ and $\beta^{M}=\left \vert \frac{1}{2}%
\ln \frac{\left \vert \vec{h}_{\operatorname{Re}}^{\prime}+\vec{h}%
_{\operatorname{Im}}^{\prime}\right \vert }{\left \vert \vec{h}%
_{\operatorname{Re}}^{\prime}-\vec{h}_{\operatorname{Im}}^{\prime}\right \vert
}\right \vert $. As a result, the eigenvalue of $\hat{H}_{0}$ is the same as
that of $\hat{H}_{\mathrm{NH}}$, i.e.,
\begin{equation}
E_{\pm}=h_{0}\pm \sqrt{\vec{h}^{2}}.
\end{equation}
For the case of $\vec{h}^{2}>0,$ $\hat{H}_{0}$\ is a Hermitian Hamiltonian,
$\hat{H}_{0}=\hat{H}_{0}^{\dagger}$, whose energy levels $E_{\pm}=h_{0}%
\pm \left \vert \vec{h}\right \vert $ are real; for the case of $\vec{h}^{2}<0,$
$\hat{H}_{0}$\ is an anti-Hermitian Hamiltonian, $\hat{H}_{0}=-\hat{H}%
_{0}^{\dagger}$, whose energy levels $E_{\pm}=h_{0}\pm i\left \vert \vec
{h}\right \vert $ are an imaginary pair.

Finally, one can see that under the non-Hermitian\ similarity transformation,
the initial basis $\{|\psi_{0}^{\mathrm{R}}\rangle \}$ is correspondingly
changed into the unique matrix basis $\{|\psi^{\mathrm{R}}\rangle \}$, i.e.,
\begin{align}
\{|\psi_{0}^{\mathrm{R}}\rangle \}  &  \rightarrow \{|\psi^{\mathrm{R}}%
\rangle \}=\{ \left \vert 1\right \rangle ^{\mathrm{R}},\left \vert 2\right \rangle
^{\mathrm{R}}\} \nonumber \\
&  =\{ \mathcal{\hat{S}}_{M}(\vec{\sigma}^{B},\beta^{B})|\psi_{0}^{\mathrm{R}%
}\rangle \} \\
&  =\{ \mathcal{\hat{S}}_{M}(\vec{\sigma}^{M},\beta^{M})\mathcal{\hat{S}}%
_{B}(\vec{\sigma}^{B},\beta^{B})|\psi_{0}\rangle \}.\nonumber
\end{align}

In addition, to describe the same non-Hermitian two-level system, one can also
use another representation based on the normal basis, $\{ \hat{H}%
_{\mathrm{NH}}^{\beta^{B}},\{|\psi_{0}\rangle \} \}$, i.e.,
\begin{equation}
\hat{H}_{\mathrm{NH}}^{\beta^{B}}=\mathcal{\hat{S}}_{B}^{-1}\hat
{H}_{\mathrm{NH}}\mathcal{\hat{S}}_{B}%
\end{equation}
and $\{|\psi_{0}\rangle \}=\{ \mathcal{\hat{S}}_{B}^{-1}|\psi_{0}^{\mathrm{R}%
}\rangle \}$. To clearly show the relationship among the three representations
for the Hamiltonians of the same non-Hermitian two-level system, $\{ \hat
{H}_{\mathrm{NH}},\{|\psi_{0}^{\mathrm{R}}\rangle \} \}$ under the initial
basis, $\{ \hat{H}_{0},\{|\psi^{R}\rangle \} \} \}$ under the matrix basis, and
$\{ \hat{H}_{\mathrm{NH}}^{\beta^{B}},\{|\psi_{0}\rangle \} \}$ under the
normal basis, we plot Fig.~2. $\mathcal{\hat{S}}_{M}$, $\mathcal{\hat{S}}_{B}%
$, and $\mathcal{\hat{S}}_{M}\mathcal{\hat{S}}_{B}$ are different
non-Hermitian similarity transformations relating these representations.

\begin{figure}[ptb]
\includegraphics[clip,width=0.62\textwidth]{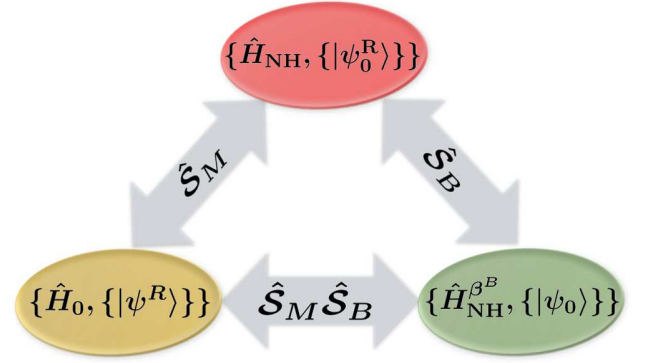}
\centering \setlength{\abovecaptionskip}{-2.cm} \vspace{-0.2cm}\caption{(Color
online) Schematic diagram of the relationship among three equivalent
representations for a two-level system: $\{ \hat{H}_{\mathrm{NH}},\{|\psi
_{0}^{\mathrm{R}}\rangle \} \}$, under the initial basis; $\{ \hat{H}%
_{0},\{|\psi^{R}\rangle \} \}$, under the matrix basis; and $\{ \hat
{H}_{\mathrm{NH}}^{\beta^{B}},\{|\psi_{0}\rangle \} \}$, under the normal
basis. Here, $\hat{H}_{\mathrm{NH}}$ is the original non-Hermitian Hamiltonian
with initial basis $\{|\psi_{0}^{\mathrm{R}}\rangle \}$. $\{|\psi^{R}%
\rangle \}=\{ \mathcal{\hat{S}}_{M}|\psi_{0}^{\mathrm{R}}\rangle \}=\{
\mathcal{\hat{S}}_{M}\mathcal{\hat{S}}_{B}|\psi_{0}\rangle \}$ is a matrix
basis that classifies different types of GEPs (M-GEPs, B-GEPs, and H-GEPs).
$\hat{H}_{\mathrm{NH}}^{\beta^{B}}=\mathcal{\hat{S}}_{B}\hat{H}_{\mathrm{NH}%
}\mathcal{\hat{S}}_{B}^{-1}$\ classifies the subclasses of B-GEPs. }%
\end{figure}

\subsection{Definition of general EPs}

We point out that the universal feature of arbitrary non-Hermitian two-level
systems with singularities is \emph{defectiveness of the matrix basis
}$\{|\psi^{\mathrm{R}}\rangle \}$\emph{ }as $\beta^{M}\rightarrow \infty$ and/or
$\beta^{B}\rightarrow \infty$\emph{ rather than energy degeneracy or state
coalescence}. Thus, we define \emph{general EPs} as follows:

\textit{Definition -- general EPs (GEPs): For an arbitrary non-Hermitian
two-level system, GEPs exist if and only if the matrix basis }$\{|\psi
^{\mathrm{R}}\rangle \}=\{ \left \vert 1\right \rangle ^{\mathrm{R}},\left \vert
2\right \rangle ^{\mathrm{R}}\}$\textit{ becomes defective, i.e., }$\frac
{1}{\mathcal{N}}^{\mathrm{R}}\langle1|1\rangle^{\mathrm{R}}=1$\textit{ and
}$^{\mathrm{R}}\langle2|2\rangle^{\mathrm{R}}=0$\ or\ $\frac{1}{\mathcal{N}%
}^{\mathrm{R}}\langle2|2\rangle^{\mathrm{R}}=1$\textit{ and }$^{\mathrm{R}%
}\langle1|1\rangle^{\mathrm{R}}=0$\textit{. Here, }$\mathcal{N}=^{\mathrm{R}%
}\langle1|1\rangle^{\mathrm{R}}+^{\mathrm{R}}\langle2|2\rangle^{\mathrm{R}}%
$\textit{ is a normalization factor. }

As a result, at GEPs, the two energy levels may not necessarily be degenerate,
i.e., $E_{+}=E_{-}$ and $E_{+}\neq E_{-}$ are both allowed; the two
eigenstates $\left \vert \tilde{\psi}_{+}^{\mathrm{R}}\right \rangle $ and
$\left \vert \tilde{\psi}_{-}^{\mathrm{R}}\right \rangle $ may not necessarily
coalesce, i.e., $\Lambda \equiv1$ and $\Lambda \equiv0$ are both allowed. In
addition, in the form of Jordan block matrix, at GEPs the algebra multiplier
of a matrix may be same to its geometric multiplier.

\subsection{Classification of general EPs}

By considering different behaviors of defective bases, we can classify GEPs
for an arbitrary non-Hermitian two-level system. Depending on the behavior of
the defective matrix basis $\{|\psi^{\mathrm{R}}\rangle \}=\{ \mathcal{\hat{S}%
}_{M}(\vec{\sigma}^{M},\beta^{M})\mathcal{\hat{S}}_{B}(\vec{\sigma}^{B}%
,\beta^{B})|\psi_{0}\rangle \},$ there are \emph{three} types of GEPs:
\emph{matrix-type GEPs} (or M-GEPs), with $\beta^{M}\rightarrow \infty$ and
$\beta^{B}\nrightarrow \infty$; \emph{basis-type GEPs} (or B-GEPs), with
$\beta^{M}\nrightarrow \infty$ and $\beta^{B}\rightarrow \infty$; and
\emph{hybrid-type GEPs} (or H-GEPs), with $\beta^{M}\rightarrow \infty$ and
$\beta^{B}\rightarrow \infty$.

In addition, there are \emph{two} different classes of B-GEPs, i.e., IB-GEPs
and IIB-GEPs. For IB-GEPs, the two eigenstates will never coalesce with each
other, i.e., $\Lambda \equiv0$; for IIB-GEPs, the two eigenstates will always
coalesce each other, i.e., $\Lambda \equiv1$.

To show the physical properties of these two classes of B-GEPs, we perform a
non-Hermitian similarity transformation $\mathcal{\hat{S}}_{B}^{-1}$ on the
initial basis $\{|\psi_{0}^{\mathrm{R}}\rangle$ and obtain a representation
under the normal basis, i.e.,
\begin{equation}
\{|\psi_{0}^{\mathrm{R}}\rangle \} \rightarrow \{ \mathcal{\hat{S}}_{B}%
^{-1}|\psi_{0}^{\mathrm{R}}\rangle \}=\{|\psi_{0}\rangle \}.
\end{equation}
Correspondingly, the original Hamiltonian $\hat{H}_{\mathrm{NH}}$ is
transformed into
\begin{equation}
\hat{H}_{\mathrm{NH}}^{\beta^{B}}=\mathcal{\hat{S}}_{B}\hat{H}_{\mathrm{NH}%
}\mathcal{\hat{S}}_{B}^{-1}.
\end{equation}
There are\emph{ two} possibilities for $\hat{H}_{\mathrm{NH}}^{\beta^{B}}$,
which correspond to the two classes of B-GEPs: one possibility is that all
elements of $\hat{H}_{\mathrm{NH}}^{\beta^{B}}$ are finite, in which case the
eigenstates do not coalesce (or $\Lambda \equiv0$), and the other is one or
more elements diverge, i.e., $(\hat{H}_{\mathrm{NH}}^{\beta^{B}}%
)_{ij}\rightarrow \infty$, in which case the eigenstates coalesce (or
$\Lambda \equiv1$). It is obvious that at IB-GEPs the algebra multiplier of the
Hamiltonian $\hat{H}_{\mathrm{NH}}$ is equal to its geometric multiplier.

Therefore, there exists a quantum phase transition between IB-GEPs (the region
without eigenstate coalescence) and IIB-GEPs (the region with eigenstate
coalescence). Let us give a simple explanation of this fact.

For IB-GEPs, the Hamiltonian $\hat{H}_{\mathrm{NH}}$ commutes with
$\vec{\sigma}^{B}$, i.e., $[\hat{H}_{\mathrm{NH}},\vec{\sigma}^{B}]=0$. As a
result, the Hamiltonian $\hat{H}_{\mathrm{NH}}$ must be written as
$\lambda \vec{\sigma}^{B}$ with $\lambda \neq0$. Under the non-Hermitian
similarity transformation $\mathcal{\hat{S}}_{B}^{-1}=e^{\beta^{B}\vec{\sigma
}^{B}}$, we have
\begin{equation}
\hat{H}_{\mathrm{NH}}^{\beta^{B}}=\mathcal{\hat{S}}_{B}^{-1}\hat
{H}_{\mathrm{NH}}\mathcal{\hat{S}}_{B}=\lambda \vec{\sigma}^{B}.
\end{equation}
Now, the matrix basis becomes normal, i.e.,
\begin{equation}
\{|\psi_{0}^{\mathrm{R}}\rangle \} \rightarrow \{ \mathcal{\hat{S}}_{B}%
^{-1}|\psi_{0}^{\mathrm{R}}\rangle \}=\{|\psi_{0}\rangle \}.
\end{equation}
Here, the $|\psi_{0}\rangle$ are eigenstates of $\vec{\sigma}^{B}$, or
$\vec{\sigma}^{B}|\psi_{0}\rangle=\pm|\psi_{0}\rangle$. Because the
eigenstates $|\psi_{\pm}^{\mathrm{R}}\rangle$ of $\hat{H}_{\mathrm{NH}%
}=\lambda \vec{\sigma}^{B}$\ are also those of $|\psi_{0}\rangle$, the state
similarity of $|\psi_{\pm}^{\mathrm{R}}\rangle$ must be zero, i.e.,
\begin{equation}
\Lambda=|\langle \tilde{\psi}_{+}^{\mathrm{R}}|\tilde{\psi}_{-}^{\mathrm{R}%
}\rangle|=0.
\end{equation}

On the other hand, for IIB-GEPs, the Hamiltonian $\hat{H}_{\mathrm{NH}}$ does
not commute with $\vec{\sigma}^{B},$ i.e., $[\hat{H}_{\mathrm{NH}},\vec
{\sigma}^{B}]\neq0$. The Hamiltonian $\hat{H}_{\mathrm{NH}}$ must be written
as
\begin{equation}
\hat{H}_{\mathrm{NH}}=\lambda \vec{\sigma}^{B}+\eta(\vec{\sigma}^{B})^{\perp}%
\end{equation}
with $\lambda^{2}+\eta^{2}\neq0$ and $\{(\vec{\sigma}^{B})^{\perp},\vec
{\sigma}^{B}\}=0$. Now, one element of $\hat{H}_{\mathrm{NH}}^{\beta^{B}%
}=\mathcal{\hat{S}}_{B}^{-1}\hat{H}_{\mathrm{NH}}\mathcal{\hat{S}}_{B}$
diverges. On the basis of $\{|\psi_{0}\rangle \}$, the divergent term is
proportional to $(%
\begin{array}
[c]{cc}%
0 & 1\\
0 & 0
\end{array}
)$ or $(%
\begin{array}
[c]{cc}%
0 & 0\\
1 & 0
\end{array}
)$. As a result, the Hamiltonian $\hat{H}_{\mathrm{NH}}^{\beta^{B}%
}=\mathcal{\hat{S}}_{B}^{-1}\hat{H}_{\mathrm{NH}}\mathcal{\hat{S}}_{B}$ is
dominated by this divergent term, and we can ignore other terms. In this case,
the state similarity of $|\psi_{\pm}^{\mathrm{R}}\rangle$ must be $1$, i.e.,
\begin{equation}
\Lambda=|\langle \tilde{\psi}_{+}^{\mathrm{R}}|\tilde{\psi}_{-}^{\mathrm{R}%
}\rangle|=1.
\end{equation}
Without sudden changing the energy levels and the defectiveness of matrix
basis, this quantum phase transition is always \textquotedblleft%
\emph{hidden}\textquotedblright.

\section{Example: 1D nonreciprocal Su--Schrieffer--Heeger model}

\subsection{The model}

In this section, we take the 1D nonreciprocal Su--Schrieffer--Heeger (SSH)
model as an example to illustrate the different types of GEPs for its
topologically protected sub-space of two edge states.

The Bloch Hamiltonian for the nonreciprocal SSH model under periodic boundary
conditions (PBC) is given by
\begin{align}
\hat{H}_{\mathrm{PBC}}(k)  &  =\sum_{k}c_{k}^{\dagger}\tau_{x}\left(
t_{1}+t_{2}\cos k\right)  c_{k}\\
&  +\sum_{k}c_{k}^{\dagger}\tau_{y}\left(  t_{2}\sin k+i\gamma \right)
c_{k}+i\varepsilon \sum_{k}c_{k}^{\dagger}\tau_{z}c_{k},\nonumber
\end{align}
where $c_{k}^{\dagger}=(c_{k,\mathrm{A}}^{\dagger},c_{k,\mathrm{B}}^{\dagger
})$; the $\tau_{i}$ are the Pauli matrices acting on the (\textrm{A} or
\textrm{B}) sublattice subspaces; $t_{1}$ and $t_{2}$ describe the intracell
and intercell hopping strengths, respectively; $\gamma$ describes unequal
intracell hopping; and $\varepsilon$ denotes the strength of an imaginary
staggered potential on the two sublattices. $t_{1}$, $t_{2}$, $\gamma$, and
$\varepsilon$ are all real. In this paper, we set $t_{2}=1$.

Under a non-Hermitian similarity transformation $\mathcal{\hat{S}%
}_{\mathrm{NHP}}$, the physics properties of the 1D nonreciprocal SSH model
under open boundary conditions (OBC) are characterized by $\hat{H}%
_{\mathrm{OBC}}(k)$ rather than $\hat{H}_{\mathrm{PBC}}(k)$\cite{Yao}. Here,
the non-Hermitian similarity transformation $\mathcal{\hat{S}}_{\mathrm{NHP}}$
given by
\begin{equation}
c_{k}^{\dagger}\rightarrow \tilde{c}_{k}^{\dagger}=c_{k-iq_{0}}^{\dagger
}=\mathcal{\hat{S}}_{\mathrm{NHP}}c_{k}^{\dagger}%
\end{equation}
or
\begin{equation}
c_{n}^{\dagger}\rightarrow \tilde{c}_{n}^{\dagger}=e^{-q_{0}(n-1)}%
c_{n}^{\dagger}=\mathcal{\hat{S}}_{\mathrm{NHP}}c_{n}^{\dagger}%
\end{equation}
where $e^{q_{0}}=\sqrt{\frac{t_{1}-\gamma}{t_{1}+\gamma}}.$ Consequently, the
effective hopping parameters become $\bar{t_{1}}=\sqrt{(t_{1}+\gamma
)(t_{1}-\gamma)}$ and $\bar{t_{2}}=t_{2}$.

To characterize the topological properties of the non-Hermitian topological
system, the non-Bloch topological invariant $\bar{w}$ of $\hat{H}%
_{\mathrm{OBC}}(k)$ is introduced, i.e.,%
\begin{equation}
\bar{w}=\frac{1}{2\pi}\int_{-\pi}^{\pi}\partial \bar{\phi}(k)dk
\end{equation}
where $\bar{\phi}(k)=\tan^{-1}(\bar{h}_{y}/\bar{h}_{x})$ and $\bar{h}_{x}%
=\bar{t}_{1}+\bar{t}_{2}\cos k$, $\bar{h}_{y}=\bar{t}_{2}\sin k$. In the
region of $\left \vert \bar{t}_{1}\right \vert <\left \vert \bar{t}%
_{2}\right \vert $ and $\bar{w}=1$, the system is a topological insulator (the
gray region in Fig. 3(a)); in the region of $\left \vert \bar{t}_{1}\right \vert
>\left \vert \bar{t}_{2}\right \vert $ and $\bar{w}=0$, the system is a normal
insulator (the white region in Fig. 3(a)). A quantum phase transition occurs
at $\left \vert \bar{t}_{1}\right \vert =\left \vert \bar{t}_{2}\right \vert $,
where the bulk energy gap under OBC is closed.

\subsection{Two-level systems from two edge states in topological phase}

In the topological phase with $\bar{w}=1$, there exist two edge states
$\left \vert \psi_{1}^{\mathrm{R}}\right \rangle $ and $\left \vert \psi
_{2}^{\mathrm{R}}\right \rangle $. These two edge states make up a
topologically protected subspace denoted by $\{ \hat{H}_{\mathrm{NH}}%
,\{|\psi_{0}^{\mathrm{R}}\rangle \} \}$. Under the biorthogonal set, the
initial basis $\{|\psi_{0}^{\mathrm{R}}\rangle \}$ is $\{ \left \vert \psi
_{1}^{\mathrm{R}}\right \rangle ,\  \left \vert \psi_{2}^{\mathrm{R}%
}\right \rangle \}$. The effective Hamiltonian $\hat{H}_{\mathrm{NH}}$ is
written as
\begin{equation}
\hat{H}_{\mathrm{NH}}=\left(
\begin{array}
[c]{cc}%
h_{11} & h_{12}\\
h_{21} & h_{22}%
\end{array}
\right)  ,
\end{equation}
where $h_{ij}=\left \langle \psi_{i}^{\mathrm{L}}\right \vert \hat{H}\left \vert
\psi_{j}^{\mathrm{R}}\right \rangle $, $i,j=1,2$. Through straightforward
calculations\cite{Ran}, we can analytically obtain the effective Hamiltonian
of the two edge states as
\begin{equation}
\hat{H}_{\mathrm{NH}}=\bar{\Delta}\sigma^{x}+i\varepsilon \sigma^{z},
\end{equation}
where $\bar{\Delta}=\frac{(\bar{t}_{2}^{2}-\bar{t}_{1}^{2})}{\bar{t}_{2}%
}(\frac{\bar{t}_{1}}{\bar{t}_{2}})^{N}$. The two energy levels for the two
eigenstates $|\psi_{+}^{\mathrm{R}}\rangle$ and $|\psi_{-}^{\mathrm{R}}%
\rangle$ are $E=\pm \sqrt{\bar{\Delta}^{2}-\varepsilon^{2}}$.

\begin{figure}[ptb]
\includegraphics[clip,width=0.63\textwidth]{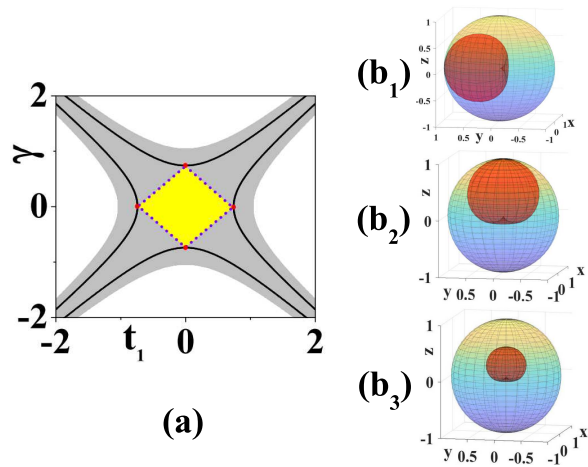}
\centering \setlength{\abovecaptionskip}{-1.cm} \vspace{-0.2cm}\caption{(Color
online) (a) is the global phase diagram for the 1D nonreciprocal SSH model
under open boundary conditions ($N=20$ and $\varepsilon=0.001$): four red dots
$\rightarrow$ M-GEPs; solid black lines $\rightarrow$ H-GEPs; yellow region
$\rightarrow$ IB-GEPs; gray region $\rightarrow$ IIB-GEPs; dashed lines
$\rightarrow$ hidden quantum phase transition between IB-GEPs and IIB-GEPs.
(b$_{1}$), (b$_{2}$) and (b$_{3}$) show examples of\ the Bloch peaches for
M-GEPs, B-GEPs, and H-GEPs, respectively. }%
\end{figure}

\subsection{General EPs}

Then, based on $\{ \hat{H}_{\mathrm{NH}},\{|\psi_{0}^{\mathrm{R}}\rangle \}
\}$, we study the GEPs for the two edge states in the 1D nonreciprocal SSH model.

In topological phase with $\bar{w}=1$ (or $\left \vert \bar{t}_{1}\right \vert
<\left \vert \bar{t}_{2}\right \vert $), we have GEPs for the two edge states
except for the Hermitian/anti-Hermitian cases at $\gamma=0$ or $t_{1}=0$. Fig.
3(a) is an illustration of the global phase diagram for GEPs: the four red
dots correspond to M-GEPs, the solid black lines correspond to H-GEPs, the
yellow region corresponds to IB-GEPs, and the gray region corresponds to
IIB-GEPs. The Bloch peaches for different types of GEPs are illustrated in
Fig. 3(b$_{1}$) (an example for M-GEPs), Fig. 3(b$_{2}$) (an example for
B-GEPs), and Fig. 3(b$_{3}$) (an example for H-GEPs).

Let us separately discuss the different types of GEPs one-by-one.

\begin{figure}[ptb]
\includegraphics[clip,width=0.63\textwidth]{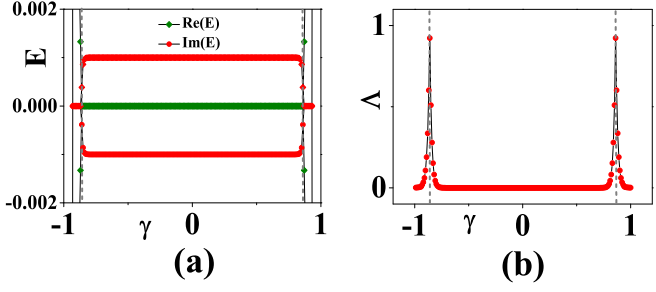}
\centering \setlength{\abovecaptionskip}{-1.cm} \vspace{-0.2cm}\caption{(Color
online) (a) is the energy levels of two edge states; (b) is the state
similarity of two edge states, $\Lambda=|\langle \tilde{\psi}_{+}^{\mathrm{R}%
}|\tilde{\psi}_{-}^{\mathrm{R}}\rangle|$. (a) and (b) are all for the case of
$N=50$ , $t_{2}=1$, $\varepsilon=0.001$ and $t_{1}=0$. The two energy levels
become degenerate and the state similarity becomes $1$ at the gray dotted line
($\gamma=\pm0.86$), where the M-GEP occurs.}%
\end{figure}

\subsubsection{M-GEPs}

Firstly, we consider the case of M-GEPs by setting $\varepsilon$ to a purely
real value and $\gamma=0$. Now, the initial basis becomes normal, i.e.,
\begin{equation}
\{|\psi_{0}^{\mathrm{R}}\rangle \}=\{|\psi_{0}\rangle \}
\end{equation}
with $\mathcal{\hat{S}}_{B}(\vec{\sigma}^{B},\beta^{B})=1$. The Hamiltonian of
the effective two-level model is reduced to
\begin{equation}
\hat{H}_{\mathrm{NH}}=i\varepsilon \sigma^{z}+\Delta_{0}\sigma^{x},
\end{equation}
where $\Delta_{0}=\frac{\left(  t_{2}^{2}-t_{1}^{2}\right)  }{t_{2}}%
(-\frac{t_{1}}{t_{2}})^{N}$.

A spontaneous $\mathcal{PT}$-symmetry-breaking transition occurs at
$\left \vert \varepsilon \right \vert =\Delta_{0}.$ The energy levels become
degenerate, shown as Fig. 4(a) i.e.,
\begin{equation}
E_{\pm}=\pm \sqrt{\varepsilon^{2}-\Delta_{0}^{2}}\rightarrow0,
\end{equation}
and the non-Hermitian similarity transformation becomes singular, i.e.,
\begin{equation}
\mathcal{\hat{S}}_{M}=e^{-\beta^{M}\cdot \sigma^{y}}%
\end{equation}
with $\beta^{M}=\left \vert \frac{1}{2}\ln \left \vert \frac{\Delta
_{0}+\varepsilon}{\Delta_{0}-\varepsilon}\right \vert \right \vert
\rightarrow \infty.$ We have an M-GEP with the following defective matrix
basis:
\begin{equation}
\{|\psi^{\mathrm{R}}\rangle \}=\{ \mathcal{\hat{S}}_{M}(\sigma^{y},\beta
^{M}\rightarrow \infty)|\psi_{0}\rangle \}.
\end{equation}
\ The state similarity $\Lambda=|\langle \tilde{\psi}_{+}^{\mathrm{R}}%
|\tilde{\psi}_{-}^{\mathrm{R}}\rangle|=1$ indicates the coalescence of the two
edge states, shown as the Fig. 4(b). From the illustration of Fig. 4, we can
obtain that at M-GEPs two energy levels become degenerate and two energy
states coalesce ($E_{+}\neq E_{-}$, $\Lambda=1$),\ shown as the gray dotted
line at $\gamma=\pm0.86$. In addition, as shown in Fig. 3(b$_{1}$), the Bloch
peach has a symmetric axis along the y-direction.

\begin{figure}[ptb]
\includegraphics[clip,width=0.63\textwidth]{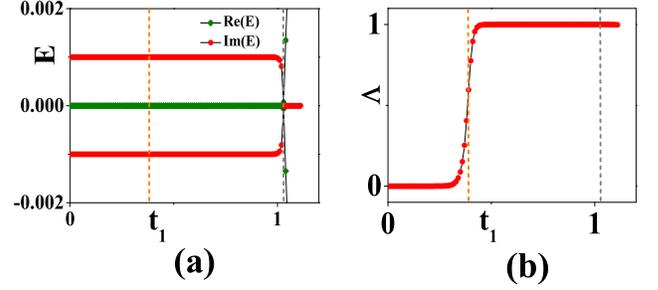}
\centering \setlength{\abovecaptionskip}{-1.cm} \vspace{-0.2cm}\caption{(Color
online) (a) is the energy levels of two edge states; (b) is the state
similarity of two edge states, $\Lambda=|\langle \tilde{\psi}_{+}^{\mathrm{R}%
}|\tilde{\psi}_{-}^{\mathrm{R}}\rangle|$. (a) and (b) are all for the case of
$N=50$, $t_{2}=1$, $\varepsilon=0.001$ and $\gamma=0.5$. The two energy levels
become degenerate ($E_{+}=E_{-}$) at the gray dotted line ($t_{1}=1.03$),
where the H-GEP occurs with $\Lambda=1$; The orange dotted line ($t_{1}=0.38$)
denoted the phase transition between IB-GEPs ($E_{+}\neq E_{-}$, $\Lambda=0$)
and IIB-GEPs ($E_{+}\neq E_{-}$, $\Lambda=1$).}%
\end{figure}

\subsubsection{B-GEPs}

Secondly, we consider the case of B-GEPs by setting $\left \vert \bar{\Delta
}\right \vert \neq \left \vert \varepsilon \right \vert $ with $t_{1}\neq0$ and
$\gamma \neq0$. Now, the initial basis becomes defective, i.e.,
\[
\{|\psi_{0}^{\mathrm{R}}\rangle \}=\{ \mathcal{\hat{S}}_{B}|\psi_{0}\rangle \}
\]
with $\mathcal{\hat{S}}_{B}=e^{-\beta^{B}\cdot \sigma_{z}}$ and $\beta
^{B}=Nq_{0}.$ Here, $q_{0}=\frac{1}{2}\ln(\frac{t_{1}-\gamma}{t_{1}+\gamma})$
is an imaginary wave vector that characterizes the non-Hermitian skin effect.
In the thermodynamic limit $N\rightarrow \infty$, we have a B-GEP with the
following defective matrix basis:
\begin{equation}
\{|\psi^{\mathrm{R}}\rangle \}=\{|\psi_{0}^{\mathrm{R}}\rangle \}=\{
\mathcal{\hat{S}}_{B}(\sigma^{z},\beta^{B}\rightarrow \infty)|\psi_{0}%
\rangle \}.
\end{equation}
In Fig. 3(a), except for the black lines and red dots, B-GEPs exist throughout
the whole topological insulator region. An interesting fact is that the energy
levels are not degenerate, shown as Fig. 5(a), i.e.,
\begin{equation}
E_{\pm}=\pm \sqrt{\bar{\Delta}^{2}-\varepsilon^{2}}\neq0.
\end{equation}
As a result, \emph{this is an example of GEPs without energy degeneracy.} Now,
the Bloch peach has a symmetric axis along the z-direction; see the
illustration in Fig. 3(b$_{2}$).

Let us discuss the classes of B-GEPs in detail.

After the application of a non-Hermitian similarity transformation
$\mathcal{\hat{S}}_{B}$, the effective two-level model $\hat{H}_{\mathrm{NH}}$
is transformed into another non-Hermitian model, i.e.,
\begin{align}
\hat{H}_{\mathrm{NH}}^{\beta^{B}}  &  =(\mathcal{\hat{S}}_{B})^{-1}\hat
{H}_{\mathrm{NH}}(\mathcal{\hat{S}}_{B})\nonumber \\
&  =\bar{\Delta}^{+}\sigma^{+}+\bar{\Delta}^{-}\sigma^{-}+i\varepsilon
\sigma^{z},
\end{align}
where $\bar{\Delta}^{+}=\bar{\Delta}\exp{(-Nq_{0})}$ and $\bar{\Delta}%
^{-}=\bar{\Delta}\exp{(Nq_{0})}$. In the thermodynamic limit $N\rightarrow
\infty$, there exist three phases: a phase with $\left \vert \bar{\Delta}%
^{+}\right \vert \rightarrow \infty$ and $\left \vert \bar{\Delta}^{-}\right \vert
\rightarrow0$, a phase with $\left \vert \bar{\Delta}^{+}\right \vert
\rightarrow0$ and $\left \vert \bar{\Delta}^{-}\right \vert \rightarrow \infty$,
and a phase with $\left \vert \bar{\Delta}^{+}\right \vert \rightarrow0$ and
$\left \vert \bar{\Delta}^{-}\right \vert \rightarrow0$.\textbf{ }At $\left \vert
\bar{\Delta}^{\pm}\right \vert =1$ or $t_{1}\pm \gamma=\pm1$, the
\textquotedblleft hidden\textquotedblright \ quantum phase transition occurs,
with a sudden change between IB-GEPs ($\Lambda=0$) and IIB-GEPs ($\Lambda=1$).
Shown as Fig. 5, the orange dotted line at $t_{1}=0.38$\ identifies the phase
transition between IB-GEPs ($E_{+}\neq E_{-}$, $\Lambda=0$) and IIB-GEPs
($E_{+}\neq E_{-}$, $\Lambda=1$). We find that the situation changes for the
state similarity $\Lambda=|\langle \tilde{\psi}_{+}^{\mathrm{R}}|\tilde{\psi
}_{-}^{\mathrm{R}}\rangle|$ of two edge states. In Fig. 6, one can see the
\textquotedblleft hidden\textquotedblright \ quantum phase transition from the
numerical results for the case of $N=20$ and $\varepsilon=0.001$ that
$\Lambda$ suddenly changes from $0$ to $1$ at $\left \vert \bar{\Delta}^{\pm
}\right \vert =1.$

\begin{figure}[ptb]
\includegraphics[clip,width=0.63\textwidth]{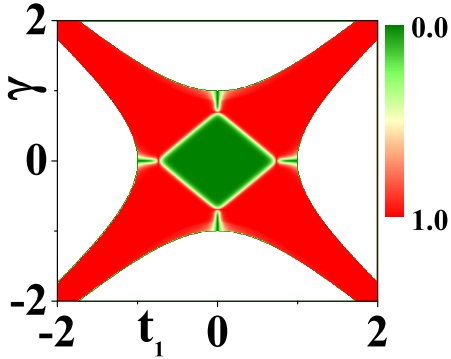}
\centering \setlength{\abovecaptionskip}{-1.cm} \vspace{-0.2cm}\caption{(Color
online) The state similarity of two edge states, $\Lambda=|\langle \tilde{\psi
}_{+}^{\mathrm{R}}|\tilde{\psi}_{-}^{\mathrm{R}}\rangle|$, for the case of
$N=20$ and $\varepsilon=0.001$ as a function of $\gamma$ and $t_{1}$. A hidden
phase transition occurs with a sudden change between IB-GEPs ($\Lambda=0$,
green regions) and IIB-GEPs ($\Lambda=1$, red regions).}%
\end{figure}

\subsubsection{H-GEPs}

Thirdly, we consider the\ case of H-GEPs by setting $\varepsilon$ to a purely
real value, $\gamma \neq0,$ and $\left \vert \bar{\Delta}\right \vert =\left \vert
\varepsilon \right \vert $. On the one hand, the initial basis is defective,
i.e.,
\begin{equation}
\{|\psi_{0}^{\mathrm{R}}\rangle \}=\{ \mathcal{\hat{S}}_{B}|\psi_{0}%
\rangle \}=\{e^{-\beta^{B}\cdot \sigma_{z}}|\psi_{0}\rangle \}
\end{equation}
with $\beta^{B}=Nq_{0}\rightarrow \infty$. In the thermodynamic limit
$N\rightarrow \infty$, the two-level system may be regarded as a B-GEP. On the
other hand, for the case of $\left \vert \frac{\bar{\Delta}}{\varepsilon
}\right \vert =1$, the same system can be regarded as an M-GEP with another
singular non-Hermitian similarity transformation $\mathcal{\hat{S}}_{B}%
(\beta^{M}\rightarrow \infty)$.

Therefore, in thermodynamic limit $N\rightarrow \infty$, we have an H-GEP with
the following defective matrix basis:
\begin{align*}
\{|\psi^{\mathrm{R}}\rangle \}  &  =\{ \mathcal{\hat{S}}_{M}(\sigma_{y}%
,\beta^{M}=\left \vert \frac{1}{2}\ln \left \vert \frac{\bar{\Delta}+\varepsilon
}{\bar{\Delta}-\varepsilon}\right \vert \right \vert \rightarrow \infty)\\
\cdot \mathcal{\hat{S}}_{B}(\sigma_{z},\beta^{B}  &  =Nq_{0}\rightarrow
\infty)|\psi_{0}\rangle \}.
\end{align*}
Now, the quantum state under the non-Hermitian similarity transformations
$\mathcal{\hat{S}}_{M}(\sigma_{y},\beta^{M})$ and $\mathcal{\hat{S}}%
_{B}(\sigma_{z},\beta^{B})$ becomes
\begin{align}
|\psi^{\mathrm{R}}\rangle &  =\frac{1}{\sqrt{2}}(\cos \frac{\theta}%
{2}+e^{-\beta^{B}}\sin \frac{\theta}{2})\left \vert 1\right \rangle \\
&  +\frac{i}{\sqrt{2}}e^{i\varphi}e^{-\beta^{M}}(\cos \frac{\theta}%
{2}-e^{-\beta^{B}}\sin \frac{\theta}{2})\left \vert 2\right \rangle \nonumber
\end{align}
(with $\theta \in \lbrack0,\pi]$ and $\varphi \in \lbrack0,2\pi]$), and the radius
of the Bloch sphere $R$\ can be obtained as
\begin{align}
R  &  =\langle \psi^{\mathrm{R}}|\psi^{\mathrm{R}}\rangle=\frac{1}{\sqrt{2}%
}[(\cos \frac{\theta}{2}+e^{-\beta^{B}}\sin \frac{\theta}{2})^{2}\\
&  +e^{-2\beta^{M}}(\cos \frac{\theta}{2}-e^{-\beta^{B}}\sin \frac{\theta}%
{2})^{2}]^{1/2}.\nonumber
\end{align}
As shown in Fig. 3(b$_{3}$), the Bloch peach, with a symmetric axis along the
z-direction, shrinks.

In addition, for the non-Hermitian SSH model, the H-GEP is an unusual
spontaneous $\mathcal{PT}$-symmetry breaking accompanied by a transition from
real to complex spectra. From Fig. 6, the two energy levels become degenerate
($E_{+}=E_{-}$) at the gray dotted line with $t_{1}=1.03$, where the H-GEP
occurs, and we can see that the state similarity of the two edge states is
constant, i.e., $\Lambda \equiv1$.

\section{Conclusion and discussion}

In this paper, we illustrate that the phenomenon of EPs in non-Hermitian
quantum systems is much more interesting than expected. For general
non-Hermitian two-level systems, there may exist three types of GEPs ---
M-GEPs (with a defective matrix basis and a normal initial basis), B-GEPs
(with a normal matrix basis and a defective initial basis), and H-GEPs (with a
defective matrix basis and a defective initial basis). In addition, there are
two classes of B-GEPs --- IB-GEPs, without eigenstate coalescence, and
IIB-GEPs, with eigenstate coalescence. By taking the topologically protected
edge states in the non-Hermitian SSH model as an example, we explore the
physical properties of GEPs. In the future, we will study higher-order GEPs in
non-Hermitian systems and attempt to develop a complete theory of GEPs.

\acknowledgments This work was supported by NSFC Grant Nos. 11974053 and 12174030.


\begin{thebibliography}{99}                                                                                               %


\bibitem {Kato}T. Kato, Perturbation Theory for Linear Operators, 2nd ed.,
Classics in Mathematics (Springer-Verlag, Berlin, Heidelberg, 1995)

\bibitem {Heiss1}W.D.Heiss, Nucl. Phys. A \textbf{144}, 417 (1970).

\bibitem {Moiseyev}N. Moiseyev and S. Friedland, Phys. Rev. A \textbf{22}, 618(1980).

\bibitem {Bender1}C. M. Bender and S. Boettcher, Phys. Rev. Lett. \textbf{80},
5243 (1998).

\bibitem {Miri}M.-A. Miri and A. Alù, Science \textbf{363}, eaar7709 (2019).

\bibitem {Minganti}F. Minganti, A. Miranowicz, R. W. Chhajlany, and F. Nori,
Phys. Rev. A\textbf{100}, 062131 (2019).

\bibitem {Bender2}C. M. Bender, D. C. Brody and H. F.Jones, Phys. Rev. Lett.
\textbf{89}, 270401 (2002); C. M. Bender, Rep. Prog. Phys.\textbf{70},
947-1018 (2007); C. M. Bender, D. C. Brody, H. F. Jones, and B. K. Meister,
Phys. Rev. Lett.\textbf{98}, 040403 (2007).

\bibitem {Heiss2}W.D.Heiss, J. Phys. A \textbf{45}, 444016 (2012).

\bibitem {Mostafazadeh}A. Mostafazadeh, Phys. Rev. Lett. \textbf{99}, 130502 (2007).

\bibitem {Lee}Y. C. Lee, M. H. Hsieh, S. T. Flammia, and R. K. Lee, Phys. Rev.
Lett. \textbf{112}, 130404 (2014).

\bibitem {Kawabata}Kohei Kawabata, Yuto Ashida, and Masahito Ueda, Phys. Rev.
Lett. \textbf{119}, 190401 (2017).

\bibitem {El-Ganainy}R. El-Ganainy, K. G. Makris, M. Khajavikhan, Z. H.
Musslimani, S. Rotter, and D. N. Christodoulides, Nat. Phys. \textbf{14}, 11 (2018).

\bibitem {Yao}S. Yao, and Z. Wang, Phys. Rev. Lett. \textbf{121}, 086803
(2018); S. Yao, F. Song, and Z. Wang, Phys. Rev. Lett. \textbf{121},136802 (2018).

\bibitem {Rotter}S. K. Özdemir, S. Rotter, F. Nori, and L. Yang, Nat.
Mater. \textbf{18}, 783 (2019).

\bibitem {Jin}S. Lin, L. Jin, and Z. Song, Phys. Rev. B. \textbf{99}, 165148 (2019).

\bibitem {Deng}T. S. Deng and W. Yi, Phys. Rev. B \textbf{100}, 035102 (2019).

\bibitem {Liu}Y. Liu, X. P. Jiang, J. Cao, and S. Chen, Phys. Rev. B
\textbf{101}, 174205 (2020).

\bibitem {Ran}X. R. Wang, C. X. Guo, and S. P. Kou, Phys. Rev. B \textbf{101},
121115(R) (2020); X. R. Wang, C. X. Guo, Q. Du, and S. P. Kou, Chin. Phys.
Lett. \textbf{37}, 117303 (2020).

\bibitem {Ashida}Y. Ashida, Z. Gong, and M. Ueda, Adv. Phys. \textbf{3}, 69 (2020).

\bibitem {Guo}C. X. Guo, X. R. Wang, C. Wang and S. P. Kou, Phys. Rev. B
\textbf{101}, 121116(R) (2020).

\bibitem {Guo1}C. X. Guo, X. R. Wang, and S. P. Kou, Europhys. Lett.
\textbf{131},27002 (2020).

\bibitem {wang}C. Wang, X. R. Wang, C. X. Guo ,and S. P. Kou,Int. J. Mod.
Phys. B \textbf{34},2050146 (2020).

\bibitem {wang1}C. Wang, M. L. Yang, C. X. Guo , X. M. Zhao, and S. P. Kou,
Europhys. Lett. \textbf{128},41001 (2019).

\bibitem {Feng}L. Feng, Z. J. Wong, R.-M. Ma, Y. Wang, and X. Zhang, Science
\textbf{346}, 972 (2014).

\bibitem {Peng}B. Peng, ¸S. K. Özdemir, S. Rotter, H. Yilmaz, M.
Liertzer, F. Monifi, C. M. Bender, F. Nori, and L. Yang, and L. Fu, Science
\textbf{346}, 328 (2014).

\bibitem {Hodaei}H. Hodaei, M.A. Miri, M. Heinrich, D. N. Christodoulides, and
M. Khajavikhan, Science \textbf{346}, 975 (2014).

\bibitem {AGuo}A. Guo, G. J. Salamo, D. Duchesne, R. Morandotti, M.
Volatier-Ravat, V. Aimez, G. A. Siviloglou, and D. N. Christodoulides, Phys.
Rev. Lett. \textbf{103}, 093902 (2009).

\bibitem {Chen}W. Chen, ¸S. K. Özdemir, G. Zhao, J. Wiersig, and
L. Yang, Nature (London) \textbf{548}, 192 (2017).

\bibitem {Hodaei1}H. Hodaei, A. U. Hassan, S. Wittek, H. Garcia-Gracia, R.
El-Ganainy, D. N. Christodoulides, and M. Khajavikhan, Nature(London)
\textbf{548}, 187 (2017).

\bibitem {El-Ganainy2}C. E. Rüter, K. G. Makris, R. El-Ganainy, D. N.
Christodoulides, M. Segev, and D. Kip, Nat. Phys. \textbf{6}, 192 (2010).

\bibitem {Schindler}J. Schindler, A. Li, M. C. Zheng, F. M. Ellis, and T.
Kottos, Phys. Rev. A \textbf{84}, 040101(R) (2011).

\bibitem {Feng2}L. Feng, M. Ayache, J. Huang, Y.-L. Xu, M.-H. Lu, Y.-F. Chen,
Y. Fainman, and A. Scherer, Science \textbf{333}, 729 (2011).

\bibitem {Regensburger}A. Regensburger, C. Bersch, M.-A. Miri, G. Onishchukov,
D. N. Christodoulides, and U. Peschel, Nature (London) \textbf{488}, 167 (2012).

\bibitem {Bittner}S. Bittner, B. Dietz, U. Günther, H. L. Harney, M.
Miski-Oglu, A. Richter, and F. Schäfer, Phys. Rev. Lett. \textbf{108},
024101 (2012).

\bibitem {MBender}C. M. Bender, B. K. Berntson, D. Parker, and E. Samuel, Am.
J. Phys. \textbf{81}, 173 (2013).

\bibitem {Hang}C. Hang, G. Huang, and V. V. Konotop, Phys. Rev. Lett.
\textbf{110}, 083604 (2013).

\bibitem {Zhu}X. Zhu, H. Ramezani, C. Shi, J. Zhu, and X. Zhang, Phys. Rev. X
\textbf{4}, 031042 (2014).

\bibitem {Brandstetter}M. Brandstetter, M. Liertzer, C. Deutsch, P. Klang, J.
Schöberl, H. E. Türeci, G. Strasser, K. Unterrainer, and S.
Rotter, Nat. Commun. \textbf{5}, 4034 (2014).

\bibitem {Popa}B.-I. Popa and S. A. Cummer, Nat. Commun. \textbf{5}, 3398 (2014).

\bibitem {Fleury}R. Fleury, D. Sounas, and A. Alù, Nat. Commun.
\textbf{6}, 5905 (2015).

\bibitem {Zhen}B. Zhen, C. W. Hsu, Y. Igarashi, L. Lu, I. Kaminer, A. Pick,
S.-L. Chua, J. D. Joannopoulos, and M. Soljačić, Nature (London)
\textbf{525}, 354 (2015).

\bibitem {Xu}H. Xu, D. Mason, L. Jiang, and J. G. E. Harris, Nature (London)
\textbf{537}, 80 (2016).

\bibitem {Peng1}P. Peng, W. Cao, C. Shen, W. Qu, J. Wen, L. Jiang, and Y.
Xiao, Nat. Phys. \textbf{12}, 1139 (2016).

\bibitem {Zhang}Z. Zhang, Y. Zhang, J. Sheng, L. Yang, M.-A. Miri, D. N.
Christodoulides, B. He, Y. Zhang, and M. Xiao, Phys. Rev. Lett. \textbf{117},
123601 (2016).

\bibitem {Assawaworrarit}S. Assawaworrarit, X. Yu, and S. Fan, Nature (London)
\textbf{546}, 387 (2017).

\bibitem {Choi}Y. Choi, C. Hahn, J. W. Yoon, and S. H. Song, Nat. Commun.
\textbf{9}, 2182 (2018).

\bibitem {Scheel}S. Scheel and A. Szameit, Europhys. Lett. \textbf{122}, 34001 (2018).

\bibitem {Wu}Y. Wu, W. Liu, J. Geng, X. Song, X. Ye, C.-K. Duan, X. Rong, and
J. Du, Science \textbf{364}, 878 (2019).

\bibitem {L. Xiao1}L. Xiao, X. Zhan, Z. H. Bian, K. K. Wang, X. Zhang, X. P.
Wang, J. Li, K. Mochizuki, D. Kim, N. Kawakami, W. Yi, H. Obuse, B. C.
Sanders, and P. Xue, Nat. Phys. \textbf{13}, 1117 (2017).

\bibitem {Naghiloo}M. Naghiloo, M. Abbasi, Y. N. Joglekar, and K. W. Murch,
Nat. Phys. \textbf{15}, 1232 (2019).

\bibitem {L. Xiao2}L. Xiao, K. Wang, X. Zhan, Z. Bian, K. Kawabata, M. Ueda,
W. Yi, and P. Xue, Phys. Rev. Lett. \textbf{123}, 230401 (2019).

\bibitem {Liu1}W. Liu, Y. Wu, C.-K. Duan, X. Rong, and J. Du, Phys. Rev. Lett.
\textbf{126}, 170506

\bibitem {ChenPX}W. C. Wang, Y. L. Zhou, H. L. Zhang, J. Zhang, M. C. Zhang,
Y. Xie, C. W. Wu, T. Chen, B. Q. Ou, W. Wu, H. Jing, and P. X. Chen, Phys.
Rev. A \textbf{103} L020201 (2021).

\bibitem {Ding}L. Ding, K. Shi, Q. Zhang, D. Shen, X. Zhang,and W. Zhang,
Phys. Rev. Lett. \textbf{126}, 083604 (2021).

\bibitem {Kane2010}M. Z. Hasan and C. L. Kane, Rev. Mod. Phys. \textbf{82},
3045 (2010).

\bibitem {Qi2011}X.-L. Qi and S.-C. Zhang, Rev. Mod. Phys. \textbf{83}, 1057 (2011).

\bibitem {Kitaev}A. Kitaev, Ann. Phys. \textbf{303}, 2 (2003);A. Kitaev, Ann.
Phys. \textbf{321}, 2(2006).

\bibitem {wen}X. G. Wen, \textit{Quantum Field Theory of Many-Body Systems,}
Oxford University Press, Oxford, (2004).

\bibitem {wen1}X. G. Wen, Int. J. Mod. Phys. B \textbf{4}, 239 (1990).

\bibitem {S. P. Kou}S. P. Kou, Phys. Rev. Lett. \textbf{102}, 120402 (2009).
J. Yu and S. P. Kou, Phys. Rev. B \textbf{80}, 075107 (2009). S. P. Kou, Phys.
Rev. A \textbf{80}, 052317 (2009).
\end{thebibliography}
\end{document}